# The Impact of the Object-Oriented Software Evolution on Software Metrics: The Iris Approach

## Ra'Fat Al-Msie'deen* and Anas H. Blasi

Department of IT, Faculty of Information Technology, Mutah University, P.O. Box 7, Mutah 61710, Karak, Jordan;
rafatalmsiedeen@mutah.edu.jo, Ablasi1@mutah.edu.jo

## Abstract

The Object-Oriented (OO) software system evolves over the time to meet the new requirements. Based on the initial release of software, the continuous modification of software code leads to software evolution. Software needs to evolve over the time to meet the new user's requirements. Software companies often develop variant software of the original one depends on customers' needs. The main hypothesis of this paper states that the software when it evolves over the time, its code continues to grow, change and become more complex. This paper proposes an automatic approach (Iris) to examine the proposed hypothesis. Originality of this approach is the exploiting of the software variants to study the impact of software evolution on the software metrics. This paper presents the results of experiments conducted on three releases of drawing shapes software, sixteen releases of rhino software, eight releases of mobile media software and ten releases of ArgoUML software. Based on the extracted software metrics, It has been found that Iris hypothesis is supported by the computed metrics.

**Keywords:** Object-Oriented Software, Software Variants, Reverse Engineering, Software Engineering, Software Evolution, Software Metrics, Software Complexity.

## 1. Introduction

Object-Oriented (OO) software systems need enhancement to be used for long time[1]. Lehman's laws[2] of software evolution show that continuous modification and development is essential to keep the software system for a long time. Moreover, Lehman has proposed that over the time, due to evolution, software system becomes complex and hard to add new features on it. The software metrics are affected by software evolution over the time[3]. In real world, software evolution could be an outcome of software maintenance[4]. This paper proposes a novel method called *Iris* to investigate the impact of the OO software evolution on software metrics. Iris stands for Impact of the object-oriented softwaRe evolutIon on Software metrics.

Lehman's laws state that during software evolution, due to growth and changes, software product becomes more complex[2]. The main objective of current approaches was to investigate the applicability of Lehman's laws of software evolution on software products using different metrics[3,4]. Iris states that during software evolution, software code becomes more complex and it continues to change and grow. Iris suggests new metrics to measure software growth and complexity.

The main *hypothesis* of this work is that, if the software system evolves over the time, the software source code becomes more complex and it may continue to grow and change. In addition, to investigate the impact of software evolution related to the code complexity and its continuous growth and change, this paper suggests using the software metrics[3] (*aka* measurements) to measure this hypothesis. Iris approach has computed the OO software metrics for several software variants. The computed metrics value for different versions has used as basis for examining the Iris hypothesis.

---

*Author for correspondence

Comparing with the *related work*, most of current approaches have designed to measure the software metrics in a single software system[2]. Moreover, there are other studies have similar approaches like what have been done in this paper[3,4]. A new suggestion that has proposed in this paper is a new software metrics that measures a software growth and complexity. In order to measure the software complexity, this paper has proposed three metrics: number of inheritance relations, number of attribute accesses and number of method invocations. To measure the growth of a software code, Iris has suggested a new metrics such as, number of identifiers, and number of public methods.

Johari and Kaur[3] proposed an approach to study the effect of software evolution on software metrics. Their study has analyzed different releases of two open source software systems (JHotDraw and Rhino) developed in Java and released via evolution processes. The key objective of the research was to examine the applicability of Lehman laws on software variants. However, Iris does not examine Lehman's laws; it is suggesting a new metrics to measure software complexity and presenting new metrics to measure software growth and change.

Kaur *et al.*[4] suggest a new method that study the applicability of Lehman laws on different releases of two software developed using C++. In their work, an experiment has been conducted on ten versions of flight gear simulator and graphics layout engine evolved over the period of eight years. Based on the extracted metrics, the laws of continuous change, growth and complexity were found applicable according to the metrics collected.

Israeli and Feitelson[6] examined the Lehman's laws in the Linux kernel. Linux kernel versions released over a period of fourteen years. Linux kernel includes 810 releases. Lehman's laws include eight laws. All laws have supported in their experiment except self-regulation and feedback system. Iris does not examine the Lehman's laws but examines the complexity, growth and change of software code during its evolution.

The current methods are designed to examine the applicability of Lehman laws on different releases of software systems. Iris studies the impact of software evolution on software metrics in a collection of software variants. During software evolution, software code becomes complex and it may continue to grow and change. However, Iris proves this hypothesis via new metrics.

The rest of the paper is organized as: section 2 shows an overview of Iris approach. Section 3 presents software metrics mining process step-by-step. Section 4 describes the experiments were conducted to validate Iris hypothesis. Finally, section 5 concludes and provides perspectives for this work.

## 2. Approach Overview

This section gives an overview of the Iris approach. It also presents software metrics used in the Iris approach and describes the toy example that illustrates the remaining of the paper.

### 2.1 Approach basics

The general objective of Iris is to study the impact of software evolution on the software metrics. Iris hypothesis states that the software during its evolution its code continues to grow, changes and becomes more complex. Iris approach proves this hypothesis via new software metrics. Iris exploits the software identifiers (*i.e.,* package, class, attribute and method) and the main code dependencies between those identifiers (*i.e.,* inheritance relation, method invocation and attribute access) to extract soft-ware metrics of a collection of software variants. Iris takes the source code of a collection of software variants as inputs and generates software metrics as output. Figure 1 shows the software metrics mining process.

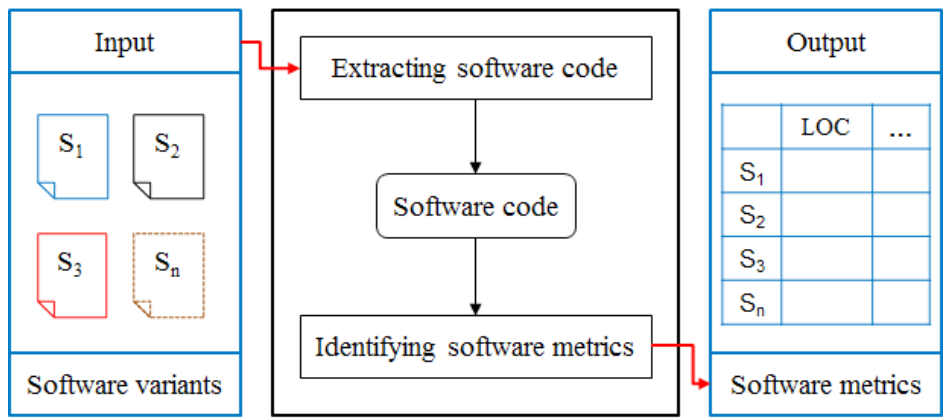

**Figure 1.** The software metrics mining process.

The first step of the software metrics mining process aims to identifying software identifiers and code dependencies based on the static code analysis[7]. The second step identifies *software-by-metric* matrix; in which the rows of the matrix correspond to software variants and the columns correspond to software metrics.

## 2.2 Metrics used in Iris approach

This section presents software metrics used in Iris approach. For measuring the software complexity at different level, it has considered inheritance relation[8] at class level, attribute access and method invocation at method level. For measuring software growth and change, it has considered software identifiers[9] to study this issue. Iris approach considers software metrics given in Table 1.

**Table 1**. Metrics used in Iris approach.

| # | Metric | Abbreviations |
|---|--------|---------------|
| 1 | Lines of code | LOC |
| 2 | Number of packages | NOP |
| 3 | Number of classes | NOC |
| 4 | Number of interfaces | NOI |
| 5 | Number of attributes | NOA |
| 6 | Number of methods | NOM |
| 7 | Number of local variables | NOL |
| 8 | Number of identifiers | NOID |
| 9 | Number of public methods | NOPM |
| 10 | Number of static methods | NOSM |
| 11 | Number of inheritance relations | NOIR |
| 12 | Number of attribute accesses | NOAA |
| 13 | Number of method invocations | NOMI |

Lines of code metric counts software lines of code, except blank lines and code comments. Inheritance relation happens when a general class (super-class) is connected to its specialized classes (sub-classes). While, attribute access occurs when methods of one class use attributes of another class. Whereas, method invocation occurs when methods of one class use methods of another class[10]. Number of identifiers metric counts all software identifiers: packages, classes, attributes and methods.

## 2.3 An Illustrative Example

As a toy example, Iris considers the drawing shapes software[11]. This software product family was developed by the authors and implemented using Java code. Drawing shapes software allows user to draw several kinds of shapes. Since its original issue in 2014, there have been several official versions so far. Table 2 summarizes changes made in each version of drawing shapes software systems. The scenarios comprise several types of changes involving include or exclude some kinds of shapes.

**Table 2.** Summary of evolution scenarios in drawing shapes software variants.

| Releases | Release description |
|----------|---------------------|
| r1 | Draw line, rectangle and oval functionality added. |
| r2 | Draw line, 3D rectangle and round rectangle functionality included. |
| r3 | Draw line, rectangle, 3D rectangle, round rectangle and oval functionality added. |

Iris does not know the software metrics in advance. Drawing shapes software variants used for better explanation of software metrics mining process. We only use the source code of software variants as input of the software metrics mining process. Figure 2 shows the first and second versions of drawing shapes variants that were developed by clone-and-own approach. Figure 2 inspired by Martinez *et al.*[12].

All releases of drawing shapes software are developed by *copy-paste-modify* technique (*i.e.,* clone-and-own approach) based on the initial release[13]. Drawing shapes variants represent a small case study (*i.e.,* third version consists of 448 lines of code).

# 3. Software Metrics Mining Process

This section presents the software metrics mining process step-by-step. According to the proposed approach, Iris identifies the software metrics in two steps as detailed in the following.

## 3.1 Extracting software code

First step of Iris approach aims to extracting the source code of software variants. Moreover, Iris approach used only the source code of software variants as inputs of the software metrics mining process. Also Iris extracts the source code of each software variant based on the static code analysis[14]. Iris code parser used to access and read the source code of software systems. Iris code parser identifies all software identifiers in addition to source code dependencies. The outputs of this step are a collection of software code files. Iris relies on the extracted code files to identify the software metrics.

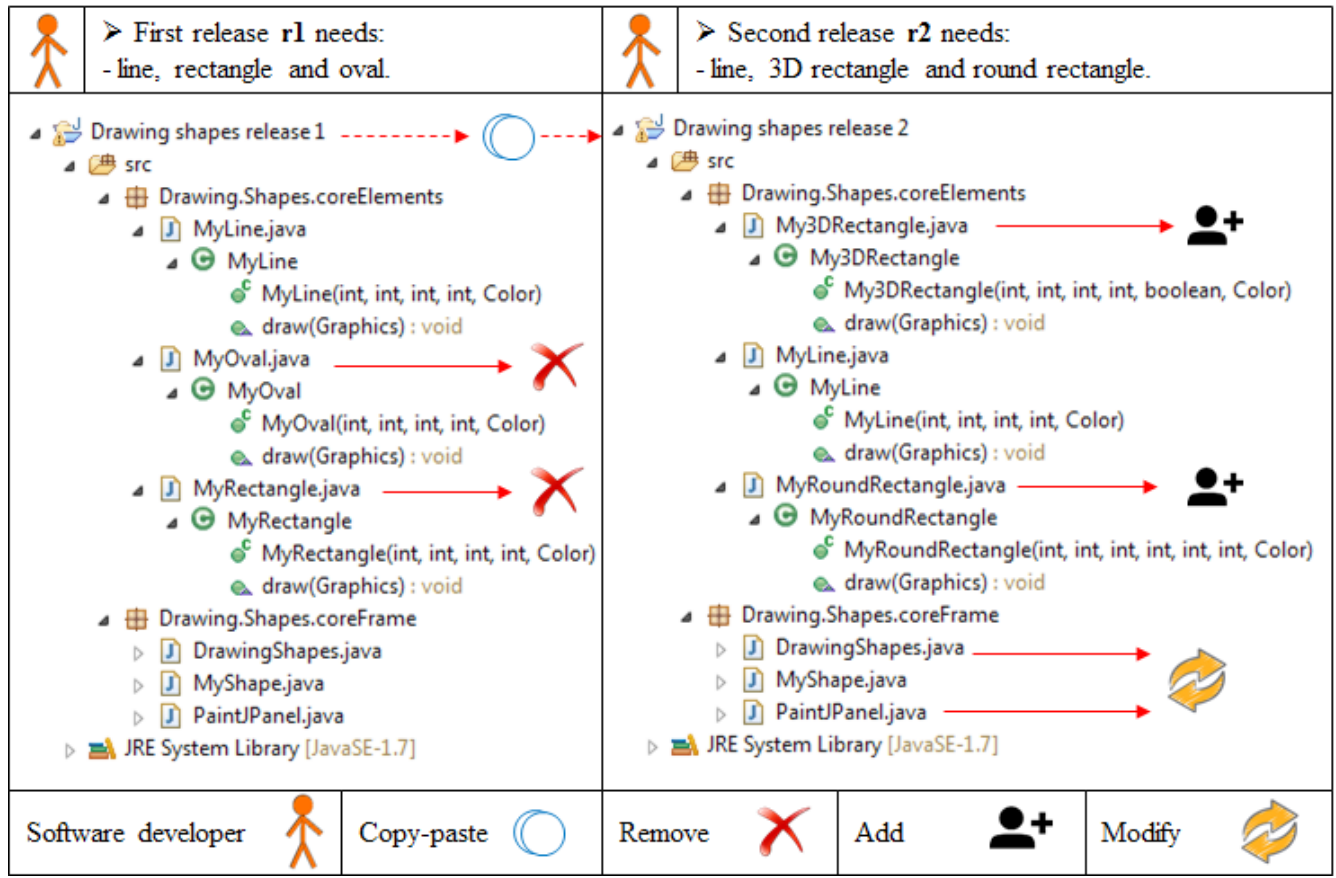

**Figure 2.** Two drawing shapes software variants developed by using clone-and-own approach.

## 3.2 Identifying software metrics

The second step of software metrics mining process aims to identifying the software metrics. This step accepts the code files and generates software metrics as output. Table 3 summarizes the obtained results for each drawing shapes software variant.

The initial version of drawing shapes software consists 386 lines of code and the newest version consists 448 lines of code (*cf.* Table 3). The size in terms of code lines has increased from the initial version to the latest version.

**Table 3.** Software metrics mined from drawing shapes software variants.

| Versions | Version date | LOC | NOP | NOC | NOA | NOM | NOID | NOI | NOIR | NOPM | NOSM | NOAA | NOMI |
|---|---|---|---|---|---|---|---|---|---|---|---|---|---|
| Drawing shapes r1 | 3/8/2014 | 386 | 4 | 6 | 16 | 29 | 55 | 0 | 5 | 1 | 26 | 125 | 99 |
| Drawing shapes r2 | 4/5/2016 | 374 | 4 | 6 | 16 | 29 | 55 | 0 | 5 | 1 | 26 | 125 | 99 |
| Drawing shapes r3 | 1/2/2018 | 448 | 4 | 8 | 16 | 33 | 61 | 0 | 7 | 1 | 30 | 161 | 139 |

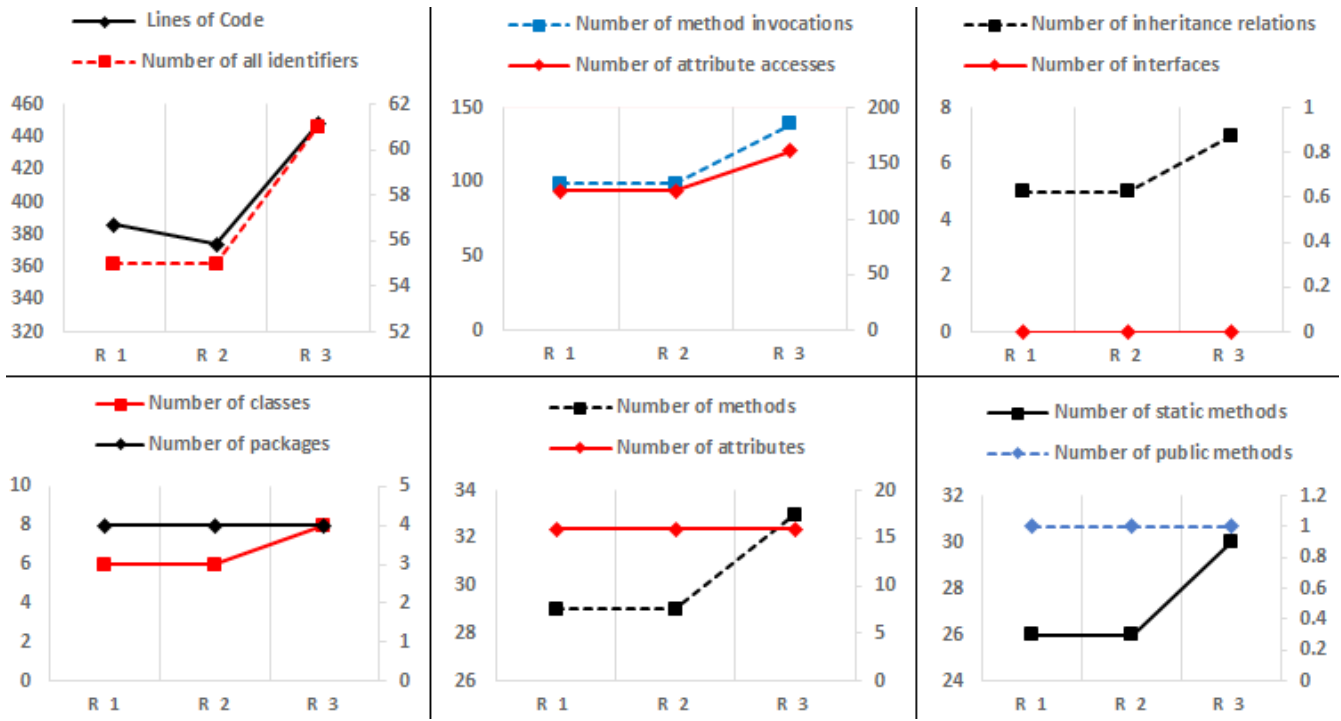

**Figure 3.** Charts summarize the extracted software metrics from drawing shapes software variants.

According to Iris hypothesis the complexity of software system tends to grow over numerous versions of software due to the software evolution process. Iris considers the main code dependencies as a metric to measure the code complexity. For example, in the newest release of drawing shapes software, the number of method invocations increased to 139. While, the number of inheritance relations increased to 7 and, at last, the number of attribute accesses increased to 161 (*cf.* Table 3).

The software code growth can interpreted as an increment in metrics size (*e.g.*, lines of code and number of packages) or increment in the software functionality (*e.g.*, number of classes and methods). Table 3 shows the growth of the size metrics and software functionalities for various versions of drawing shapes software. For example, lines of code increased from 386 to 448 in the newest version of drawing shapes. Also, the number of methods increased from 29 to 33 in the latest release of drawing shapes software. In Figure 3, charts summarize the extracted software metrics from drawing shapes software variants.

It is hard to distinguish between code growth and change. The change of software code can be due to some activities such as: software maintenance or added new functionality to the software. In case of drawing shapes software, version r1 and r2 are absolutely the same in terms of number of classes and methods, but with different functionalities. Finally, the calculated metrics prove that software code becomes more complex and continues to grow and change during software evolution.

## 4. Experimentation

This section presents the experiments that conducted to validate Iris hypothesis. It also presents the case studies and, at last, it presents the extracted software metrics.

To validate Iris proposal, experiments conducted on four Java open-source software systems: drawing shapes, rhino, mobile media and ArgoUML. The four case studies show different sizes: ArgoUML (large system), rhino (medium system), mobile media (small system) and drawing shapes (small system). The four case studies are well documented as well.

Rhino[15] software is an open-source application of JavaScript written completely in Java. Moreover, Rhino is embedded into Java applications to offer scripting to end users. Since its original issue in 1999, there have been sixteen versions so far. Table 4 presents the mined software metrics of all Rhino software releases using Iris prototype[16].

**Table 4.** Software metrics mined from Rhino software variants.

| Releases | Release date | LOC | NOP | NOC | NOA | NOM | NOID | NOI | NOIR | NOPM | NOSM | NOAA | NOMI |
|---|---|---|---|---|---|---|---|---|---|---|---|---|---|
| rhi. 1.4R3 | 10/5/1999 | 20335 | 7 | 95 | 955 | 1156 | 2213 | 16 | 21 | 324 | 862 | 19566 | 5031 |
| rhi. 1.5R1 | 10/9/2000 | 29495 | 10 | 137 | 1186 | 1621 | 2954 | 16 | 33 | 455 | 1127 | 29084 | 8708 |
| rhi. 1.5R2 | 27/7/2001 | 32060 | 7 | 131 | 1442 | 1760 | 3340 | 18 | 45 | 418 | 1035 | 32648 | 8945 |
| rhi. 1.5R3 | 27/1/2002 | 32421 | 8 | 130 | 1455 | 1777 | 3370 | 17 | 45 | 426 | 1025 | 32766 | 8932 |
| rhi. 1.5R4.1 | 10/2/2003 | 33800 | 8 | 142 | 1474 | 1891 | 3515 | 19 | 48 | 469 | 1011 | 35270 | 8957 |
| rhi. 1.5R4 | 21/4/2003 | 33718 | 8 | 142 | 1472 | 1884 | 3506 | 19 | 48 | 468 | 1007 | 35235 | 8946 |
| rhi. 1.5R5 | 25/3/2004 | 36051 | 8 | 145 | 1532 | 1944 | 3629 | 18 | 48 | 559 | 1023 | 36698 | 8806 |
| rhi. 1.6R1 | 29/11/2004 | 37961 | 10 | 144 | 1621 | 2027 | 3802 | 18 | 33 | 603 | 1054 | 37059 | 9331 |
| rhi. 1.6R2 | 19/9/2005 | 37771 | 12 | 141 | 1662 | 2021 | 3836 | 18 | 30 | 603 | 1027 | 36748 | 9215 |
| rhi. 1.6R3 | 24/7/2006 | 37946 | 12 | 141 | 1672 | 2032 | 3857 | 18 | 30 | 606 | 1033 | 36947 | 9244 |
| rhi. 1.6R4 | 10/9/2006 | 37960 | 12 | 141 | 1672 | 2033 | 3858 | 18 | 30 | 606 | 1033 | 36961 | 9250 |
| rhi. 1.6R5 | 19/11/2006 | 37960 | 12 | 141 | 1672 | 2033 | 3858 | 18 | 30 | 606 | 1033 | 36961 | 9250 |
| rhi. 1.6R6 | 30/7/2007 | 39914 | 13 | 152 | 1723 | 2124 | 4012 | 19 | 34 | 630 | 1077 | 38828 | 9777 |
| rhi. 1.6R7 | 20/8/2007 | 39930 | 13 | 152 | 1723 | 2125 | 4013 | 19 | 34 | 630 | 1078 | 38837 | 9785 |
| rhi. 1.7R1 | 6/3/2008 | 42830 | 13 | 165 | 1841 | 2270 | 4289 | 20 | 36 | 659 | 1151 | 41882 | 10718 |
| rhi. 1.7R2 | 22/3/2009 | 43425 | 11 | 167 | 1854 | 2301 | 4333 | 21 | 36 | 669 | 1176 | 42227 | 10763 |

Mobile media[17] application is Java-based open source software that manipulates image, music and video on mobile phones. Since its original issue was in 2007, there have been eight official versions so far. Mobile media application has been built as a software product line in eight versions. Figueiredo et al.[18] summarize the evolution (aka changes made) in each release. Table 5 presents the mined software metrics from mobile media software variants.

ArgoUML[19] software is a Java-based application. It is open source software and used for designing systems in the unified modelling language. The original ArgoUML software consists of nine features (aka functionalities) such as: class diagram. Since its initial version was in 2010, there have been several releases so far. Couto et al.[20] summarizes the evolution scenarios in ArgoUML software variants. The scenarios comprise several types of changes involving disable and enable some features. Table 6 presents the mined software metrics from ArgoUML software variants. Iris performed an evaluation of the execution time (in ms) of its algorithm using the ArgoUML software variants. The algorithm execution time is equal to 12843 ms.

The extracted software metrics have proven the Iris hypothesis. During software evolution, therefore, the software code becomes more complex and the software code continues to grow and change.

# 5. Conclusion and future work directions

This paper proposed Iris approach that study the impact of software evolution on software metrics. Furthermore, the hypothesis of this approach stated that the software during the evolution processes, its code continued to grow, changed and became more complex. Iris approach was implemented on numerous case studies. Based on the results obtained, the extracted metrics proved the validity of the Iris hypothesis.

**Table 5.** Software metrics mined from mobile media software variants.

| Releases | Release date | LOC | NOP | NOC | NOA | NOM | NOID | NOI | NOIR | NOPM | NOSM | NOAA | NOMI |
|---|---|---|---|---|---|---|---|---|---|---|---|---|---|
| Mobile media r1 | 3/9/2007 | 760 | 10 | 16 | 56 | 92 | 174 | 1 | 8 | 0 | 84 | 631 | 299 |
| Mobile media r2 | 3/9/2007 | 1050 | 15 | 25 | 62 | 124 | 226 | 1 | 17 | 0 | 116 | 731 | 347 |
| Mobile media r3 | 3/9/2007 | 1197 | 15 | 26 | 71 | 140 | 252 | 1 | 17 | 0 | 129 | 842 | 415 |
| Mobile media r4 | 3/9/2007 | 1246 | 15 | 26 | 74 | 143 | 258 | 1 | 17 | 0 | 132 | 898 | 446 |
| Mobile media r5 | 3/9/2007 | 1387 | 15 | 31 | 75 | 160 | 281 | 1 | 22 | 1 | 147 | 922 | 585 |
| Mobile media r6 | 3/9/2007 | 1823 | 17 | 38 | 106 | 200 | 361 | 1 | 26 | 1 | 187 | 1246 | 781 |
| Mobile media r7 | 3/9/2007 | 2214 | 17 | 47 | 133 | 239 | 436 | 1 | 35 | 1 | 212 | 1530 | 959 |
| Mobile media r8 | 3/9/2007 | 2645 | 17 | 51 | 166 | 271 | 505 | 1 | 39 | 1 | 247 | 1790 | 1200 |

**Table 6.** Software metrics mined from ArgoUML software variants.

| Releases | Release date | LOC | NOP | NOC | NOA | NOM | NOID | NOI | NOIR | NOPM | NOSM | NOAA | NOMI |
|---|---|---|---|---|---|---|---|---|---|---|---|---|---|
| Ar1 | 4/6/2010 | 120348 | 90 | 1939 | 3977 | 14904 | 20910 | 156 | 1402 | 654 | 11518 | 75813 | 56758 |
| Ar2 | 4/6/2010 | 82924 | 63 | 1510 | 2783 | 11934 | 16290 | 150 | 1040 | 584 | 9248 | 54771 | 40537 |
| Ar3 | 4/6/2010 | 118189 | 90 | 1939 | 3716 | 14901 | 20646 | 156 | 1402 | 653 | 11517 | 73749 | 55330 |
| Ar4 | 4/6/2010 | 104029 | 77 | 1718 | 3496 | 13567 | 18858 | 150 | 1226 | 595 | 10308 | 67423 | 50361 |
| Ar5 | 4/6/2010 | 114969 | 86 | 1881 | 3831 | 14483 | 20281 | 156 | 1351 | 650 | 11230 | 72517 | 54544 |
| Ar6 | 4/6/2010 | 117636 | 87 | 1898 | 3906 | 14684 | 20575 | 156 | 1363 | 653 | 11392 | 74483 | 55421 |
| Ar7 | 4/6/2010 | 117201 | 88 | 1906 | 3913 | 14667 | 20574 | 156 | 1371 | 654 | 11351 | 74190 | 55529 |
| Ar8 | 4/6/2010 | 118769 | 88 | 1920 | 3948 | 14783 | 20739 | 156 | 1385 | 654 | 11426 | 75050 | 56117 |
| Ar9 | 4/6/2010 | 116431 | 90 | 1904 | 3861 | 14508 | 20363 | 156 | 1371 | 652 | 11246 | 73596 | 55146 |
| Ar10 | 4/6/2010 | 118066 | 89 | 1921 | 3900 | 14710 | 20620 | 156 | 1387 | 652 | 11397 | 74355 | 55771 |

For future work, Iris approach would include more metrics such as coupling between classes. Moreover, Iris approach would study more open source software systems. Also, Iris plans to exploit formal concept analysis[21] to show the impact of software evolution on software metric values using the AOC-poset[22]. Iris will use formal concept analysis to identify the common metric values (*i.e.*, no changes made) and different metric values (*i.e.*, changes made) across software variants. For example, in the drawing shapes software, all variants have the same number of packages and different lines of code.

# 6. References

1. Drouin N, Badri M, Touré F. Analyzing Software Quality Evolution using Metrics: An Empirical Study on Open Source Software. Journal of software. 2013; 8(10): 2462-2473.

2. Lehman M. Laws of Software Evolution Revisited. In: proceedings of the 5th European Workshop on Software Process Technology. Springer-Verlag; 1996. P. 108-124.

3. Johari K, Kaur A. Effect of Software Evolution on Software Metrics: An Open Source Case Study. ACM SIGSOFT Software Engineering Notes. 2011; 36(5): 0163-5948.

4. Kaur T, Ratti N, Kaur P. Applicability of Lehman Laws on Open Source Evolution: A Case study. International Journal of Computer Applications. 2014; 93(18): 0975-8887.

5. Chidamber S, Chris K. A Metrics Suite for Object Oriented Design. IEEE Transactions on Software Engineering. 1994; 20(6): 476-493.

6. Israeli A, Feitelson G. The Linux Kernel as a Case Study in Software Evolution. Journal of Systems and Software. 2010; 83(3): 485-501.

7. Al-Msie'deen R. Visualizing object-oriented software for understanding and documentation. International Journal of Computer Science and Information Security. 2015; 13(5): 18-27.

8. Al-Msie'deen R, Seriai A, Huchard M, Urtado C, Vauttier S. Mining features from the object-oriented source code of software variants by combining lexical and structural similarity. In: proceedings of the IEEE 14th International Conference on Information Reuse & Integration. IEEE Computer Society; 2013. P. 586-593.

9. Al-Msie'deen R. Automatic labeling of the object-oriented source code: The Lotus Approach. Science International-Lahore "Sci.Int.(Lahore)". 2018; 30(1): 45-48.

10. Al-Msie'deen R. Reverse Engineering Feature Models from Software Variants to Build Software Product Lines: REVPLINE Approach [PhD thesis]. University of Montpellier 2; 2014.

11. Drawing shapes software variants: https://sites.google.com/site/ralmsideen/tools

12. Martinez J, Ziadi T, Bissyandé T, Klein J, Traon Y. Automating the Extraction of Model-Based Software Product Lines from Model Variants. In: proceedings of the 30th IEEE/ACM International Conference on Automated Software Engineering. IEEE; 2015. P. 396-406.

13. Al-Msie'deen R, Seriai A, Huchard M. Reengineering Software Product Variants into Software Product Line: REVPLINE Approach. LAP LAMBERT Academic Publishing; 2014.

14. Al-Msie'deen R, Seriai A, Huchard M, Urtado C, Vauttier S, Eyal Salman H. Feature Location in a Collection of Software Product Variants Using Formal Concept Analysis. In: proceedings of the Safe and Secure Software Reuse - 13th International Conference on Software Reuse. Springer; 2013. P. 302-307.

15. Rhino home page: https://developer.mozilla.org/en-US/docs/Mozilla/Projects/Rhino

16. Iris prototype: https://sites.google.com/site/ralmsideen/tools

17. Mobile media home page: http://homepages.dcc.ufmg.br/~figueiredo/spl/icse08/

18. Figueiredo E, Cacho N, Sant'Anna C, Monteiro Mario, Kulesza U, Garcia A, Soares S, Ferrari F, Khan S, Filho F, Dantas F. Evolving software product lines with aspects: an empirical study on design stability. In: proceedings of the 30th International Conference on Software Engineering. ACM; 2008. P. 261-270.

19. ArgoUML-SPL website: http://argouml-spl.tigris.org/

20. Couto M, Valente M, Figueiredo E. Extracting Software Product Lines: A Case Study Using Conditional Compilation. In: proceedings of the 15th European Conference on Software Maintenance and Reengineering. IEEE Computer Society; 2011. P. 191-200.

21. Al-Msie'deen R, Huchard M, Seriai A, Urtado C, Vauttier S. Reverse Engineering Feature Models from Software Configurations using Formal Concept Analysis. In: proceedings of the Eleventh International Conference on Concept Lattices and Their Applications. CEUR-WS.org; 2014. P. 95-106.

22. Al-Msie'deen R, Huchard M, Seriai A, Urtado C, Vauttier S, Al-Khlifat A. Concept Lattices: A Representation Space to Structure Software Variability. In: proceedings of the 5th International Conference on Information and Communication Systems. IEEE; 2014. P. 1-6.